%% file: br_align.tex
\documentclass[preprint,11pt]{elsarticle}
\journal{Journal of Systems Architecture}
\usepackage[left=2.5cm,right=2.5cm,top=2.5cm,bottom=2.5cm]{geometry} 
\usepackage{ucs}
\usepackage[T1]{fontenc}
\usepackage[utf8x]{inputenc}
\usepackage{amsmath}
\usepackage{hyperref} %
\usepackage{array}
\usepackage{breakurl} %
\usepackage{pslatex}
\newcolumntype{L}[1]{>{\raggedright\let\newline\\\arraybackslash\hspace{0pt}}m{#1}}
\newcolumntype{C}[1]{>{\centering\let\newline\\\arraybackslash\hspace{0pt}}m{#1}}
\newcolumntype{R}[1]{>{\raggedleft\let\newline\\\arraybackslash\hspace{0pt}}m{#1}}
\usepackage{multirow}
\usepackage{multicol}
\usepackage{verbatim}
\usepackage{graphicx}
\usepackage{cprotect}
\graphicspath{{./rysunki/}{./rysunki.src/}}

\include{wz_commands}

\begin{document}
\sloppy
\begin{frontmatter}
\title{Automatic latency balancing in VHDL-implemented complex pipelined systems}
\author{Wojciech M. Zabołotny\corref{a1}}
\cortext[a1]{Tel. +48 22 234 7717; fax.: +48 22 825 2300}
\ead{wzab@ise.pw.edu.pl}
\address{Institute of Electronic Systems, Warsaw University of Technology,
 ul. Nowowiejska 15/19, 00-665 Warszawa, Poland}
 \begin{abstract}
 Balancing (equalization) of latency in parallel paths in the pipelined data processing system is an important problem.
 If those paths delay data by different numbers of clock cycles, the data arriving at the processing blocks 
 are not properly aligned in time, and incorrect results are produced. Manual correction of latencies is
 a tedious and error-prone work. This paper presents an automatic method of
 latency equalization in systems described in VHDL.
 The method is based on simulation and is portable between different simulation
 and synthesis tools. The method does not increase the complexity of the synthesized design
 comparing to the solution based on manual latency adjustment.
 The example implementation of the proposed methodology together with a simple design
 demonstrating its use is available as an open source project under BSD license.
 \end{abstract}
\end{frontmatter}
 \section{Introduction}
 The pipeline architecture is known for a very long time and used to increase
 the throughput of digital blocks\wzcite{Hallin1972880}. This concept has been also
 early adopted to signal or data processing systems implemented in FPGA\wzcite{614780}.
 The pipeline architecture allows to increase the clock frequency, because the complex
 operations, that would result in long critical paths in FPGA are divided into multiple
 significantly simpler operations. Those operations may be performed 
 in a single clock cycle even at the much higher clock frequency.
 The time needed to process the set of data will be the same or even slightly 
 longer, due to the introduction of additional registers.
 However, the overall throughput of such system will increase because in each clock cycle,
 the system can accept a new set of data, and results of processing of certain previous
 data set are produced on the output.
 Of course such a system will introduce a latency of certain number of clock cycles,
 between the moment of delivery of the data set to the input and the moment when results 
 of its processing are available on the output.

\begin{figure}[t]
 {   
 \begin{center}
    \begin{tabular}{c}
    \includegraphics[width=0.9\linewidth]{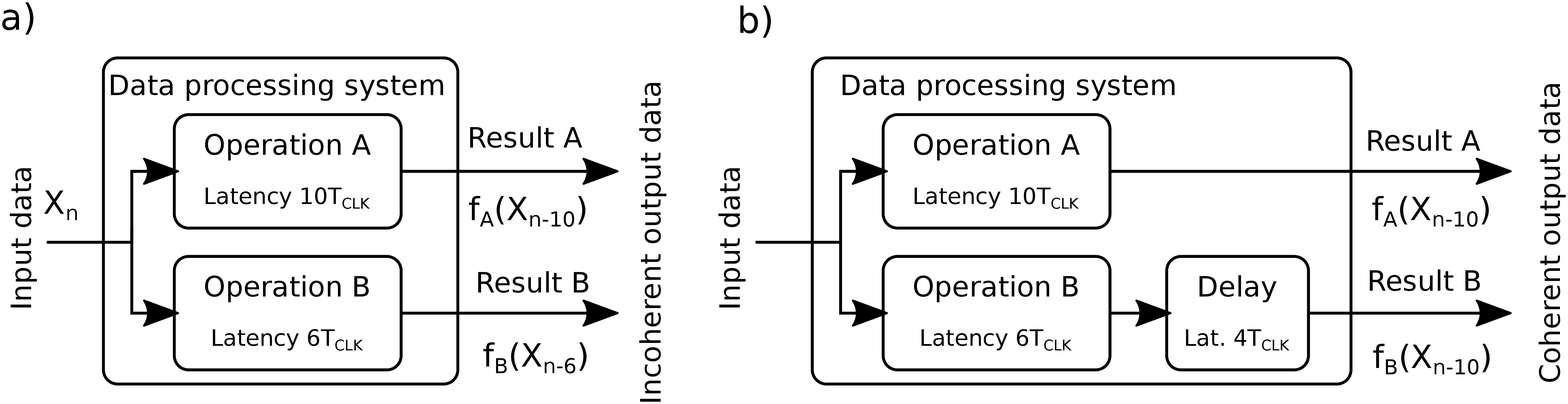}
    \end{tabular}
    \end{center}
    \caption
    { \label{fig:pipeline-demo1}
     An example of the data processing system performing two pipelined operations
     in parallel.
     a) The output data are incoherent: results of operation B are produced four clock
     periods (4~$T_{CLK}$) before results of the operation A.
     b) To assure output data coherency it was necessary to add the 4~$T_{CLK}$ delay 
     after the operation B block.
    }
 }
 \end{figure}

 Implementation of algorithms in pipelined architecture is more complicated when    
 the processing consists of different operations performed in parallel,
 and each of them requires a different number of elementary
 single-cycle operations. 
 The latency of each operation is different, and if we want to produce coherent results 
 on the output, we need to add certain delay block (typically consisting of shift registers)
 at the output of the faster block (see Figure~\ref{fig:pipeline-demo1}).
 
\begin{figure}[t]
 {   
 \begin{center}
    \begin{tabular}{c}
    \includegraphics[width=0.9\linewidth]{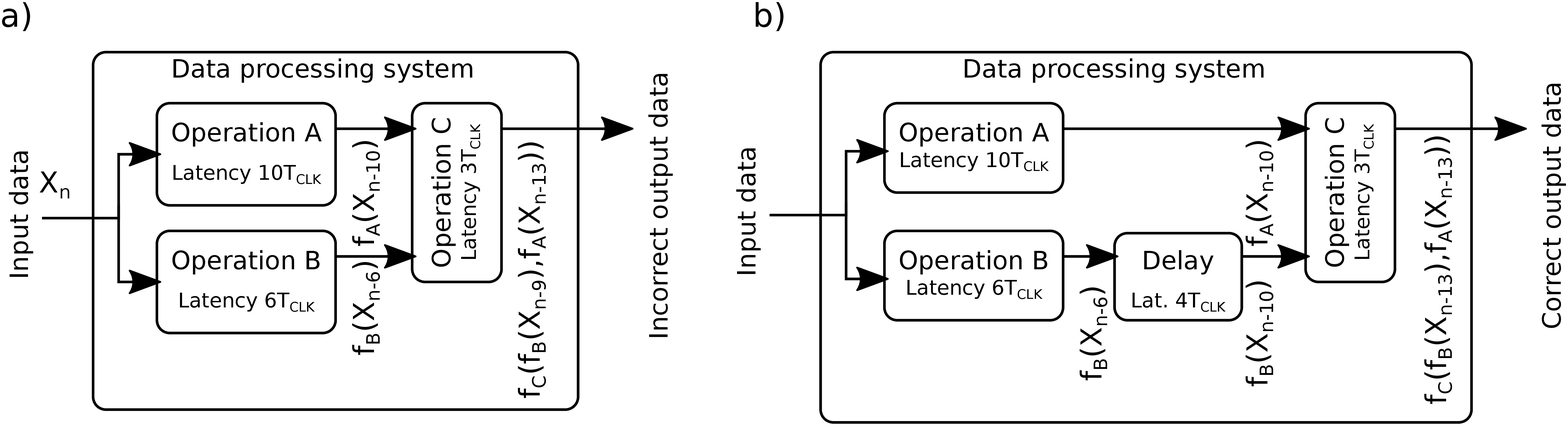}
    \end{tabular}
    \end{center}
    \caption
    { \label{fig:pipeline-demo2}
     An example of the data processing system where results of two operations calculated with 
     different latencies are used as arguments for a third operation.
     a) Without additional delays, the output data are incorrect, as arguments for operation C were calculated from different data sets.
     b) To assure correct output data, it was necessary to add the 4~$T_{CLK}$ delay 
     after the operation B block. Now both arguments of operation C are derived from the same
     data set.
    }
 }
 \end{figure}

 The problem is even more significant when results of such operations are used by another
 processing block. In this case, the results will be incorrect because operands 
 of the final operation are calculated from input data originating from
 different datasets (see Figure~\ref{fig:pipeline-demo2}).
 Again the solution is to equalize the latencies by introducing the appropriate delay block.
  In real applications, the system may contain multiple paths with different latencies,
  which must be equalized to ensure the proper operation.
  A practical example of such system may be the Overlap Muon Track Finder trigger
  for CMS detector at CERN\wzcite{doi:10.1117/12.2073380, omtf-poster}, where
  multiple data processing operations are performed in parallel, in pipelined blocks
  to produce the muon candidates\footnote{In fact, the problems with proper synchronization
  of data paths in this design were an inspiration to intensify work on the methodology
  described in this paper.}.
  Calculation of latencies in different, often quite complex paths may be a tedious task.
  Unfortunately, this work often has to be repeated, as the latencies may change during 
  the development of the system. 
  The latencies may vary due to modifications of the algorithm, but their change may be
  also enforced by modification of other parts of the IP core.
  For example when occupancy of the FPGA increases (e.g. due to addition of other blocks)
  the routing of signal becomes more complicated, and to achieve timing closure it may 
  be necessary to increase the number of pipeline 
  stages\wzcite{xlx-ufast-des-met-guide,xlx-timing-closure-user-guide, alt-timing-closure-met}.
  Therefore such designs with manually handcrafted latencies equalization are difficult 
  to maintain and reuse, and a method providing automatic delay
  balancing is needed.
  The proposed method should not impair portability of the design. Therefore, 
  it should be compatible with a possibly broad range of design tools.
  Particularly, the method should be applicable for systems implemented entirely
  in VHDL.
 \section{Available solutions}
 Of course, the problem of latency equalization between paths in
 pipelined designs is not new.
 The graphical tools, allowing to build data or signal processing systems from 
 predefined blocks implementing basic operations addressed that problem more than 14 years ago.
 Old versions of Xilinx System Generator for Simulink provided
 the ``sync'' block, which operation is described as follows:
 {\em ``The Xilinx Sync Block synchronizes two to four channels of data so that
their first valid data samples appear aligned in time with the outputs.
The input of each channel is passed through a delay line and then
presented at the output port for that channel. The lengths of the delay
lines embedded in this block, however, are adaptively chosen at the
start of simulation so that the first valid input samples are aligned.
Thus, no data appears on any channel until a first valid sample has been received into
each channel.''} (\wzcite{xlx-xsg-2.1}, page 47)\label{sec:xlx-xsg-2.1}. 
 This sync block was later synthesized using the hardware shift registers
(\wzcite{xlx-acc-dsp-designs-using-fpgas}, slide 22).

 The modern block-based tools also provide similar functionality.
 For example, the Altera DSP Builder can automatically add delays in paths
 with lower latency ``to ensure that all the input data reaches each functional
 unit in the same cycle''\wzcite{dsp-bld-adv-blkset, url-eetimes-fpga-tool}.
 No detailed information about this methodology, revealing the implementation details
 is disclosed, though.

 The article~\wzcite{6861592} describes the system level retiming,
 automatic pipelining and delay balancing (including the multi-rate pipelining) implemented
 in the MathWorks HDL Coder~\wzcite{del-bal-hdl-coder}. The delay balancing algorithm 
 used by the authors depends on the transformation of the design
 into the Parallel Implementation Representation (PIR), and further analysis of the PIR graph.
 There are no known tools able to convert the generic HDL code into the PIR form, and, therefore,
 this solution is not suitable for designs implemented in pure VHDL.

 Finally, the existing solutions have significant disadvantages:
 \begin{itemize}
 \item They are available only for systems built in graphical environments from predefined blocks 
 (however the user may also add his or her own block with needed functionality).
 \item They are closed solutions offered for the particular proprietary environment. Therefore, they are not portable.
 \item Due to their closed source nature, it is not clear how the latency balancing is implemented and if it
 can be reused in designed entirely based on HDL description. The only exceptions are:
  \begin{itemize}
  \item The old ``Xilinx Sync Block'' which uses the approach based on simulation, where the main concept is described in the accompanying documentation. The interesting thing, however, is that this block has been removed from newer versions
 of Xilinx System Generator (see \wzcite{xlx-xsg-9.1}, page 6);
  \item The algorithm implemented in the MathWorks HDL Coder that unfortunately utilizes a special intermediate
  representation of the design. This representation may be created from the Simulink model, but not from the arbitrary VHDL source.
 \end{itemize}
 \end{itemize}
 As we can see from the above review. If we want solution applicable on the level of VHDL source,
 a new approach is necessary.

 \section{Latency analysis and equalization in VHDL based designs}
 The VHDL source fully describes the behavior of the system. Therefore, one could
 try to find the data path latencies via analysis of the source code.
 Unfortunately, calculation of latency introduced in the VHDL code may be extremely difficult.
 For example, a single sequential process may introduce either one clock period
($T_{CLK}$) latency (see Figure~\ref{fig:hdl-latencies-1} a) or a latency of a few clock
    periods (see Figure~\ref{fig:hdl-latencies-1} b).
\begin{figure}
 \begin{minipage}{\linewidth}
 {
 \scriptsize
 \begin{multicols}{2}
	\small
    a)\\
	\scriptsize
\begin{verbatim}
[...]
signal sig_in, sig_out : std_logic;
[...]
p1: process (clk) is
begin  -- process p1
  if clk'event and clk = '1' then    
    if rst_p = '1' then              
      sig_out <= '0';
    else
      sig_out <= sig_in;
    end if;
  end if;
end process p1;
[...]
\end{verbatim}
	\vfill
	\columnbreak
	\small
    b)\\
	\scriptsize
\begin{verbatim}
[...]
signal sig_in, sig_out : std_logic;
signal sig_del1,sig_del2: std_logic;
[...]
p1: process (clk) is
begin  -- process p1
  if clk'event and clk = '1' then 
    if rst_p = '1' then           
      sig_out <= '0';
      sig_del1 <= '0';
      sig_del2 <= '0';
    else
      sig_del1 <= sig_in;
      sig_del2 <= sig_del1;
      sig_out <= sig_del2;
    end if;
  end if;
end process p1;
[...]
\end{verbatim}
 \end{multicols}
 }
 \end{minipage}
 \vspace{3mm}
 \caption{\label{fig:hdl-latencies-1}
 A simple process introducing different latencies between sig\_in and sig\_out signals. 
 a) The process introduces the latency of $1 T_{CLK}$. b) The process introduces the latency of  $3 T_{CLK}$ 
 }
\end{figure}    
When the structural description is used, the latency results from the serial connection
of blocks introducing their latencies. As those blocks may be implemented in other files,
it would be necessary to find a method to propagate information about introduced latencies
from the file, which defines the particular block to the another one, in which this block
is instantiated. 
The situation is even more complicated when we consider ``if generate'' and ``for generate'' 
statements.
The final conclusion is that it is impossible to derive different data path latencies
from the VHDL code without duplicating significant part of a functionality of the VHDL 
compiler. Maybe it could be possible to add such feature to an open source compiler
like GHDL\wzcite{url-ghdl}, but this is out of the scope of this paper, as it is obviously not
a portable solution.

\section{Simulation-based analysis and equalization of latencies - simplified approach}
An important part of the development of IP cores for FPGA is a preparation of 
testbenches allowing to verify correct operation of the design in simulation.
Therefore, a method allowing to check and equalize data path latencies in a simulation using
a dedicated testbench may be an acceptable solution.
Such approach was employed by the ``Xilinx Sync Block'', mentioned in section~\ref{sec:xlx-xsg-2.1}.
However, it seems that the method based only on the time of arrival of first valid data 
may be not fully reliable. It is desirable that the latency of different paths is checked or 
verified during the whole simulation period.

The general idea of the proposed method is to supplement (in simulation only) each data set with the time marker (TM) describing the moment (the clock period number), in which these data were delivered to the analyzed system.
Therefore in simulation the system must be equipped with an additional block, generating the current TM value.
In the simplest implementation the TM may be just an integer signal, starting from certain value (e.g. -1)
and increased every clock pulse\label{sec:time-markers-1st} (more detailed description of time markers
implementation is available in Section~\ref{sec:time-markers-2nd}).
	
During the processing, the time markers should travel through the system together 
with the data and results of their processing.

Of course, the time markers, and all logic associated with their processing must be included
only in simulation. Particularly, they should not increase the complexity of the synthesized design.
Fortunately, most synthesis tools offer the ``\verb|--pragma translate_off|'' and ``\verb|--pragma translate_on|'' metacomments allowing to exclude certain fragments of the VHDL code from synthesis.
Using them, we can ensure that only the delay blocks, needed to balance latencies in parallel 
paths will be added to the synthesized design.

\begin{figure}[t]
		 {   
		 \begin{center}
		    \begin{tabular}{c}
		    \includegraphics[width=0.85\linewidth]{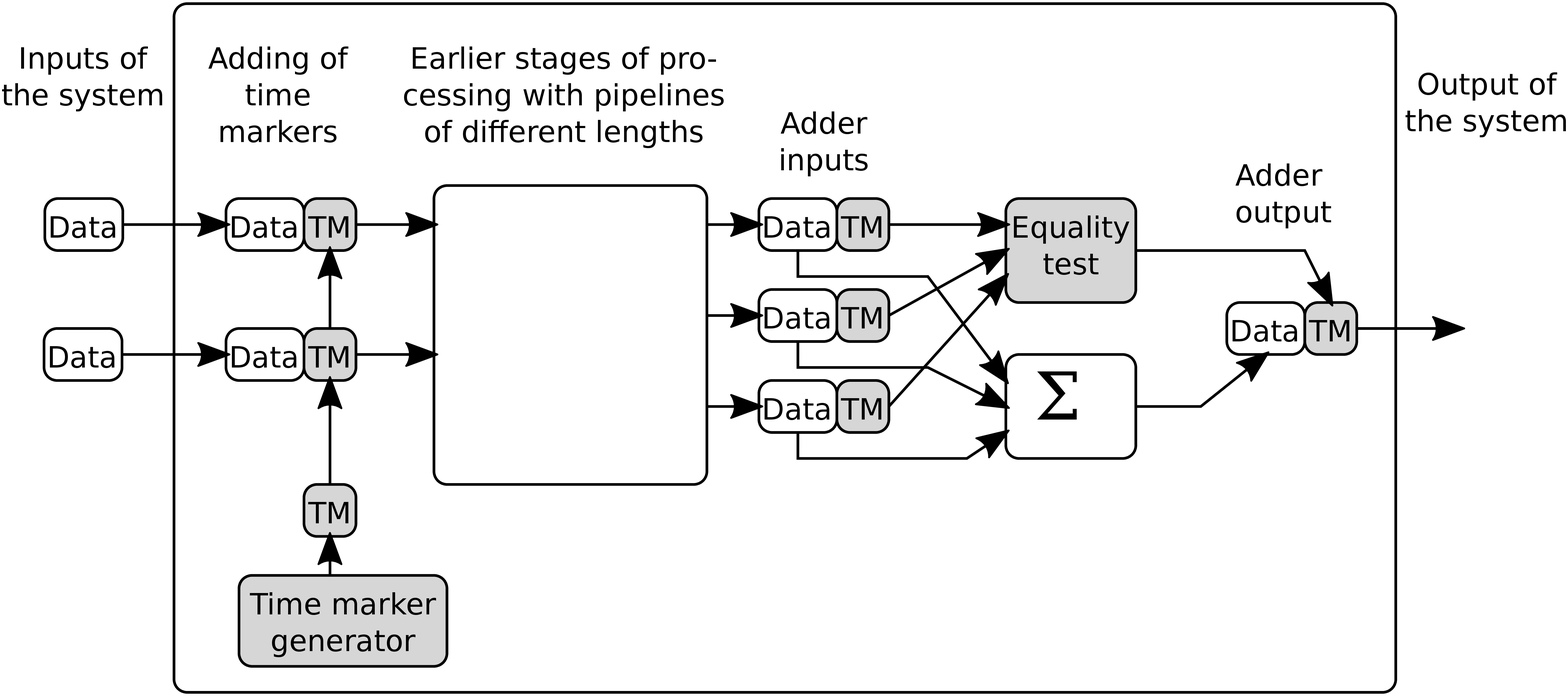}
		    \end{tabular}
		    \end{center}
		    \cprotect\caption
		    { \label{fig:adder-block1}
		     An adder as an example of the processing block implementing the described method.
		     
		     The data entering the system in the simulation are labeled with the time markers.
		     The preprocessing of data involves pipelines with different latency.
		     Finally, the sum of the three values resulting from the preprocessing is calculated.
		     Equality of time markers on the adder inputs is verified, and the same time marker
		     is produced on the output.
		     
		     Grayed objects are used only for simulation. They are excluded from synthesis using the
		      ``\verb|--pragma translate_off|'' and ``\verb|--pragma translate_on|'' metacomments.
		    }
		 }
		 \end{figure}

An example of an adder block implementing the described method is shown in Figure~\ref{fig:adder-block1}.
	
\begin{figure}[t]
  {   
   \begin{center}
    \begin{tabular}{c}
    \includegraphics[width=0.9\linewidth]{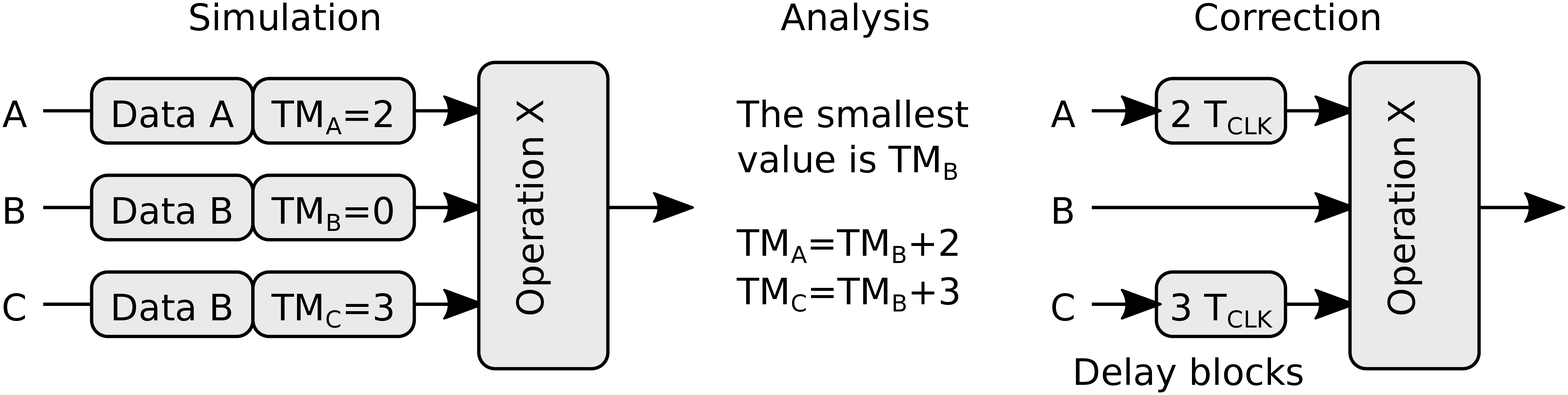}
    \end{tabular}
    \end{center}
    \caption
    { \label{fig:correction1-3inputs}
     The idea of simulation based latency correction. During the simulation, in a certain place of the 
	 design three data sets: A, B and C are used to perform operation X. The time markers (TM) associated
	 with the data sets are compared and found to be different. Therefore in the data paths with
	 the highest values of TM the delay blocks are added. The latency of added delay is equal to the
	 difference between the minimal TM and the TM in the particular path.
	}
  }
 \end{figure}

Whenever an operation on two or more subsets of data is performed, the time markers should
be checked, and in case if they are different, it is a symptom of unequal data path
latencies. In such case, the simulation should be aborted. The difference between the
time markers should be written to the file, and used to correct the design.
The shorter data path (or data paths if more than two data subsets were used as operands)
should be supplemented with additional delay block with latency equal to the detected difference (see Figure~\ref{fig:correction1-3inputs}).
Such a process should be performed iteratively until the design is found to work properly. Unfortunately, this method may require multiple iterations because 
each simulation-analysis-correction cycle allows to correct latencies on the input of one block only.

The preferable method should provide equalization of all latencies in a single
simulation-analysis-correction cycle.

\section{Simulation-based latency analysis and correction - improved approach}
In the first approach, the simulation was stopped, when the first inconsistency of time markers was detected.
However most pipelined systems may work even with misaligned data.
Of course, the results will be incorrect, but the system may be further simulated, and time markers differences between input data in other blocks may be analyzed.
There is, however, one problem. If the input time markers are equal, the time marker
of the result is simply copied from them. However, what should be the output time marker if the input time markers are not equal?
To allow proper analysis of latencies in the rest of the system, we should imitate the appropriate latency equalization. The latency equalization is achieved by introducing the delay
blocks, which results in the decrease of the time marker.
Therefore to imitate the proper equalization, the output time marker should be set to the
lowest one from the input time markers.
Of course, such situation must be reported, as the processing results will be incorrect because 
data are not properly aligned in time.
Additionally also the values of input time markers must be somehow reported, as they will be used to find the proper latencies of delay blocks in the correction phase.
	
\begin{figure}[t]
 {   
 \begin{center}
   \begin{tabular}{c}
    \includegraphics[width=0.5\linewidth]{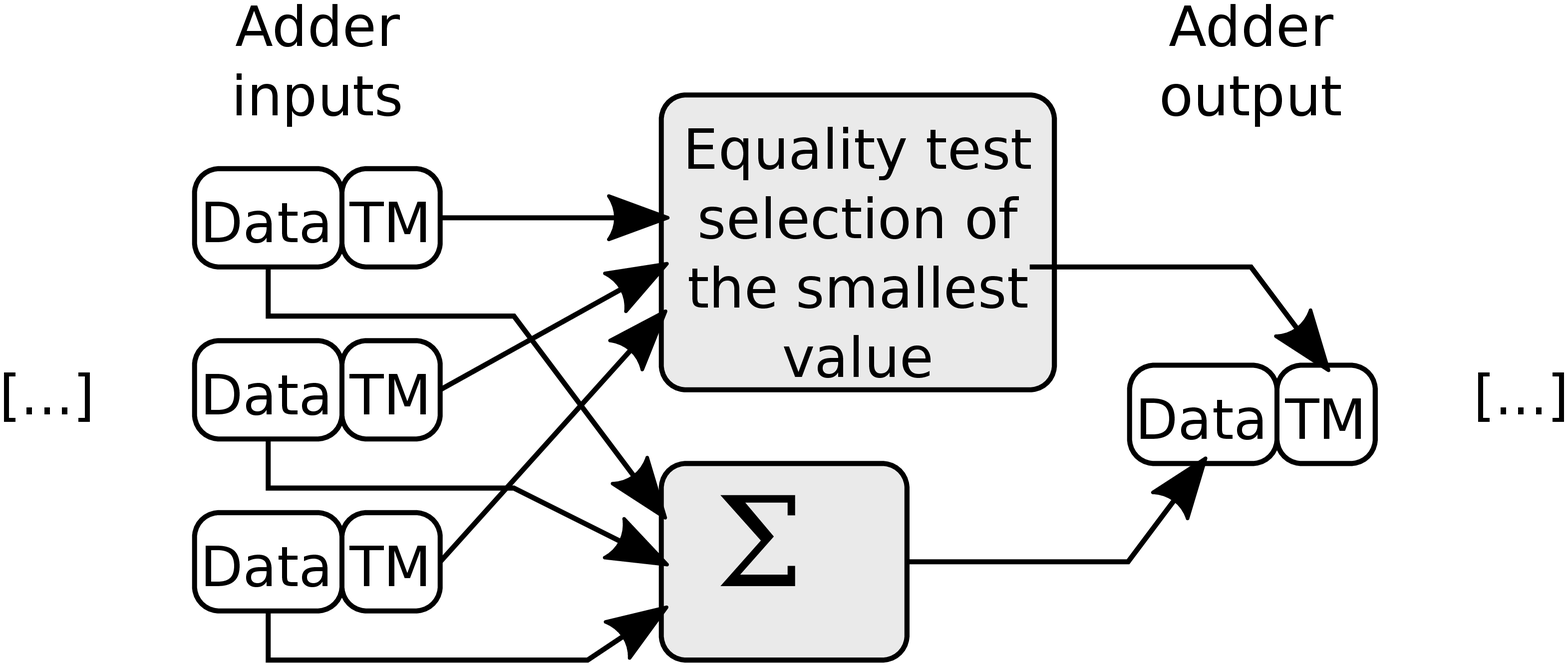}
   \end{tabular}
 \end{center}
 \caption
 { \label{fig:adder-block2}
   An adder as an example of the processing block implementing the improved method.
   A sum of the three input values is calculated. Equality of time markers on the input
   is verified. If they are equal, the same time marker is produced on the output.
   In case of inequality, we pretend that latencies are properly equalized. As we can only increase the delay by adding delay blocks, the minimal time marker is produced on the output.
   In that way, from the point of view of time markers, we can imitate the operation of the system after proper latency balancing in this block (of course the data are still misaligned, and results are incorrect).
   }
 }
\end{figure}
	
Using the described method we can test latencies of all paths in the system and calculate delays of all necessary delay blocks in a single simulation-analysis-correction cycle.
	
Certainly the testbench should also allow to test the properly latency-balanced design at the end. Therefore it must offer two modes of operation:
\begin{itemize}
 \item {\bf The analysis mode}, in which the time marker inequalities do not cause the simulation failure and in each block the output time marker is set to the smallest one from the input time markers 
 \item {\bf The final test mode}, in which any difference between time markers causes the simulation error
\end{itemize}

\section{Implementation of the proposed method in VHDL}
To allow inclusion of the time marker in the processed data, those data should be encapsulated in a record type, with optional (used only in simulation) time marker field. An example of code implementing such a record type and the adder using this type is shown in Figure~\ref{fig:adder-impl-example}.
	\begin{figure}
	\begin{minipage}{\linewidth}
	{
	\scriptsize
	\begin{multicols}{2}
\begin{verbatim}
[...]
-- pragma translate_off
subtype T_TIME_MARKER is integer;
-- pragma translate_on

-- User data without time marker
subtype T_USER_DATA is unsigned(17 downto 0);

type T_USER_DATA_MRK is record
    data : T_USER_DATA;
    -- pragma translate_off
    tm  : T_TIME_MARKER;
  -- pragma translate_on
end record T_POS_INT_MRK;
[...]
entity adder_tm is
  port (
    in1  : in  T_USER_DATA_MRK;
    in2  : in  T_USER_DATA_MRK;
    in3  : in  T_USER_DATA_MRK;
    out1 : out T_USER_DATA_MRK)
end entity adder_tm;

architecture example of adder_tm is

begin

  out1 <= in1 + in2 + in3 ;
  -- pragma translate_off
  -- We check delays only on active edge 
  -- of the clock! (otherwise there will be
  -- a lot of false alarms due do delta cycles)
  tm1 : process(clk) is
    variable out_tm : T_TIME_MARKER;
  begin
    if not ( in1.tm = in2.tm and 
             in2.tm = in3.tm) then
       -- We assume, that we have a special function
       -- for reporting unequal time markers
       report_inequality(in1.tm, in2.tm, in3.tm);
    end if
    out_tm := in1.tm;
    if in2.tm < out_tm then
      out_tm := in2.tm;
    end if; 
    if in3.tm < out_tm then
      out_tm := in3.tm;
    end if; 
    out1.tm <= out_tm;
  end process tm1;
  -- pragma translate_on
  
end architecture example;

\end{verbatim}
	\end{multicols}
	}
	\end{minipage}
	\vspace{3mm}
	\caption{\label{fig:adder-impl-example}
	An example definition of the type encapsulating the user data and the time marker and of the adder using this type.}
	\end{figure}
If the user had to modify all his or her processing blocks to include the TM handling (as in Figures~\ref{fig:adder-block1} and~\ref{fig:adder-block2}), the proposed method would be very inconvenient.
To simplify its adoption, the dedicated ``Latency Checking and Equalizing'' blocks (LCEQ) are introduced, 
The LCEQ block should offer configurable number of signal paths and should behave in the following way:
\begin{itemize}
\item In the analysis mode:
\begin{itemize}
\item Checks the time markers on its input, reporting all detected inequalities, additionally the time markers values should be recorded for further analysis.
\item Verifies the time markers on its output (after delay blocks) and in the case of their 
inequality, copies the smallest time marker to all outputs (to allow single-cycle analysis, as described previously).
\end{itemize}
\item In the final test mode: 
\begin{itemize}
\item Checks the time markers on its outputs and abort the simulation in the case of any inequality.
\end{itemize}
\end{itemize}

\begin{figure}[t]
 {   
 \begin{center}
    \begin{tabular}{c}
	    \includegraphics[width=0.95\linewidth]{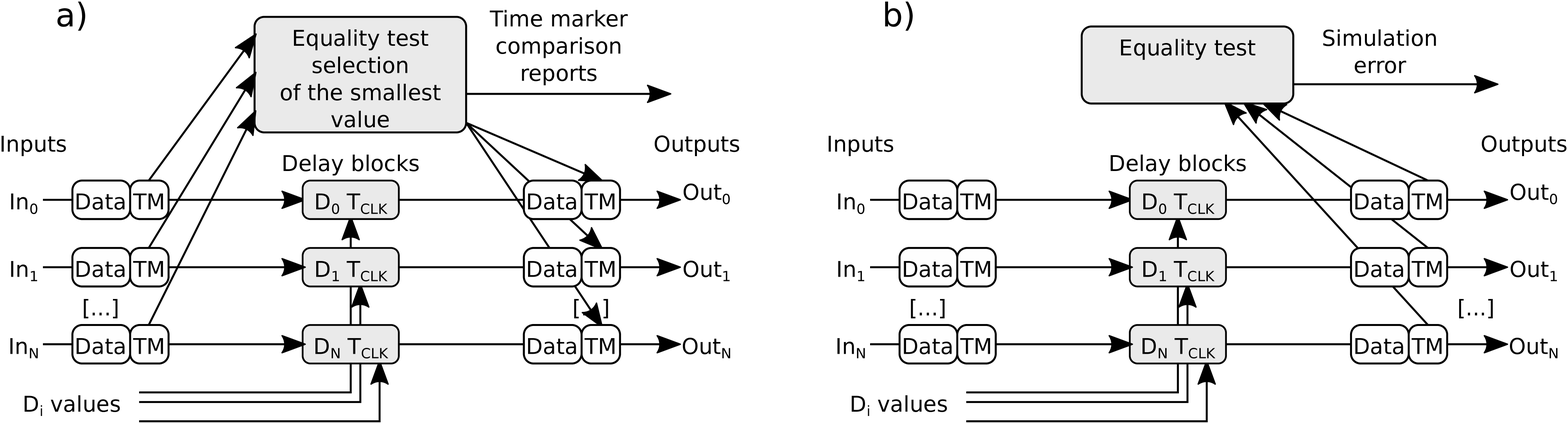}
    \end{tabular}
  \end{center}
 \cprotect\caption
  { \label{fig:lat-check-eq-1}
  The block diagram of the proposed latency checking and equalizing block.
  a) The block works in the analysis mode. b) The block works in the final test mode.
  }
 }
\end{figure}

Additionally it should be possible to configure the latency value introduced by the LCEQ in each path. The block diagram of the proposed LCEQ block is shown in Figure~\ref{fig:lat-check-eq-1}.

A possible implementation of such block in VHDL is shown in Figure~\ref{fig:lateq-sample-impl1}.
\begin{figure}
 \begin{minipage}{\linewidth}
  {
   \scriptsize
   \begin{multicols}{2}
\begin{verbatim}
library ieee;
use ieee.std_logic_1164.all;
use ieee.numeric_std.all;
library work;
use work.lateq_pkg.all;
use work.ex1_pkg.all;
use work.lateq_read_pkg.all;

entity lateq is
  generic (
    LEQ_ID : string  := "X";
    NCHANS : integer := 2
    );
  port (
    -- groups of inputs and outputs 
    din   : in  T_USER_DATA_SET(0 to NCHANS-1);
    dout  : out T_USER_DATA_SET(0 to NCHANS-1);
    -- system ports
    clk   : in  std_logic;
    rst_p : in  std_logic
    );
end lateq;

architecture beh of lateq is
  type T_DELAY is array (integer range <>) of T_USER_DATA_MRK;
  signal s_out : T_USER_DATA_SET(0 to NCHANS-1);
begin
  g1 : for i in 0 to NCHANS-1 generate
    constant NDEL : integer := lateq_read_delays(LEQ_ID, i);
    signal delay  : T_DELAY(0 to NDEL) := 
	          (others => C_USER_DATA_MRK_INIT);
  begin
    s_out(i) <= delay(0);
    -- handle the case, where latency is above 0
    gp0 : if NDEL > 0 generate
      pd0 : process(clk, rst_p) is
      begin
        if clk'event and clk = '1' then
          if rst_p = '1' then
            for i in 0 to NDEL-1 loop
              delay(i) <= C_USER_DATA_MRK_INIT;
            end loop;
          else
            delay(NDEL-1) <= din(i);
            for i in 1 to NDEL-1 loop
              delay(i-1) <= delay(i);
            end loop;
          end if;
        end if;
      end process pd0;
    end generate gp0;
    -- handle the case where latency is 0 
    -- (just copy input to output)
    gn0 : if NDEL = 0 generate
      delay(0) <= din(i);
    end generate gn0;
  end generate g1;

  -- Reporting of delays only in alayzis mode
  ig1 : if C_LATEQ_MODE = 0 generate
    --pragma translate off
    pc1 : process(clk, rst_p) is
    begin
      if clk'event and clk = '1' then
        if rst_p = '1' then
          null;
        else
          for i in 0 to NCHANS-1 loop
            lateq_report_delay(LEQ_ID, i, din(i).lateq_mrk);
          end loop;
          lateq_report_end(LEQ_ID);
        end if;
      end if;
    end process pc1;
    --pragma translate on
  end generate ig1;

  -- Aborting the simulation in final verification mode
  ig2 : if C_LATEQ_MODE = 1 generate
    --pragma translate off
    pc2 : process(clk, rst_p) is
      variable latm : T_LATEQ_MRK;
    begin
      if clk'event and clk = '1' then
        if rst_p = '1' then
          null;
        else
          latm := s_out(0).lateq_mrk;
          for i in 1 to NCHANS-1 loop
            if lateq_mrk_cmp(latm,s_out(i).lateq_mrk) /= 0 then
              report "ERROR: Inequal latencies in block " & LEQ_ID &
                " chan 0:" & lateq_mrk_to_str(latm) & " chan " &
                integer'image(i) & ":" & 
                lateq_mrk_to_str(s_out(i).lateq_mrk)
                severity FAILURE;
            end if;
          end loop;
        end if;
      end if;
    end process pc2;
    --pragma translate on
  end generate ig2;

  -- The process, which assigns outputs  
    pu : process(s_out) is
      --pragma translate off
      variable dmin : T_LATEQ_MRK;
     --pragma translate on
    begin
      --pragma translate off
      dmin := s_out(0).lateq_mrk;
      --pragma translate on
      dout <= s_out;
      --pragma translate off
      for i in 0 to NCHANS-1 loop
        if lateq_mrk_cmp(dmin, s_out(i).lateq_mrk) > 0 then
          dmin := s_out(i).lateq_mrk;
        end if;
      end loop;
      -- now we have found the dmin, so set it in all outputs
      for i in 0 to NCHANS-1 loop
        dout(i).lateq_mrk <= dmin;
      end loop;
      --pragma translate on
    end process pu;

end beh;
\end{verbatim}
   \end{multicols}
  }
 \end{minipage}
 \vspace{1mm}
 \cprotect\caption
 {\label{fig:lateq-sample-impl1}
 A sample implementation of the latency checking and equalizing block.
 The number of equalized paths is configured with the NCHANS generic parameter.
 Values of input time markers in each clock cycle are reported by the 
 \verb|lateq_report_delay| function. The end of each set is marked by the
 \verb|lateq_report_end| function.
 The latency of the delay block in each path is defined by the \verb|lateq_read_delays|
 function.
 Comparison of time markers is performed by the \verb|lateq_mrk_cmp| function.
 }
\end{figure}
 
Presented implementation of the LCEQ block may be used only when all paths carry
data of the same type. It can be acceptable in some applications, but to assure 
maximal flexibility it should be possible to define the type of data in each path independently.
Unfortunately, the VHDL language supported by most simulation and synthesis 
tools does not allow to implement a port that is an array of records of different types. The VHDL-2008\wzcite{Ashenden:1261178} introduces generic types, but even with that it still does not provide the necessary functionality. We must also consider the fact that 
VHDL-2008 is still not fully supported by most simulation and synthesis tools.
Therefore for such more general case with different types of data, another solution
is necessary.
Instead of providing the fully versatile block, there is a tool, which generates the dedicated LCEQ block for given number of paths and given types of data. 
\label{sec:lateqgen-1st}
More details are provided in section~\ref{sec:lceq-generation}.
\section{Practical implementation of the proposed method}
\label{sec:practical-implementation}
The first, ``proof of the concept'' implementation of the proposed method has been implemented and is available under open source BSD license on the OpenCores
website\wzcite{url-opencores-lateq}.
This project contains two implementations of the proposed method.
The first one uses one type for all processed data (directory hdl\_single\_type), and the 
second one uses different types for different processed data (directory 
hdl\_various\_types).

Both implementations use the same sample data processing system (described in section~\ref{sec:example-system}) as a demonstration
example.
\subsection{Generation of time markers}
\label{sec:time-markers-2nd}
As it was already mentioned in section~\ref{sec:time-markers-1st}, in the simplest implementation, 
one can use just integer numbers as time markers.
For example, the -1 value may be set as the initial value for all time markers, and
then time markers for input data are generated starting from 0 and increasing by 1 after each clock pulse.
That allows special handling of uninitialized blocks.
Such implementation has, however, one significant disadvantage. After $2^{31}$ clock pulses,
the time marker value will achieve the maximum value and an attempt to increase it will generate a simulation error.
For longer simulations another approach is needed, in which after reaching the maximum value, the time marker will return to 0. Of course, that solution needs the modified
implementations of comparison and subtraction of time marker values, to handle the ``wrapped'' values properly.

\subsection{Reporting of time markers}
The essential part of the proposed methodology is reporting of time markers from different inputs in LCEQ blocks and delivering them to the program that calculates
latencies of necessary delay blocks.
In the tested implementation, the time markers are simply written to the file. In each clock pulse the value from each input of each LCEQ block is written to the file
in a line containing:
\begin{itemize}
\item the unique identifier (LEQ\_ID) of the particular LCEQ block
\item the number of the input
\item the value of the time marker.
\end{itemize}
After the markers from each input in that clock cycle are reported, yet another
line containing only the LEQ\_ID and the word "end" is written to the file.
Such solution allows the analysis tool to check if the latency difference remains
constant during the whole simulation.
In the future implementations, it may be possible to connect the analysis tool directly 
to the VHDL simulator via named sockets or VHPI interface. That will eliminate
writing a huge amount of data to the disk. Additionally the parallel operation of the simulator and analysis tool may reduce the execution time of the simulation-analysis-correction cycle on a multiprocessor machine.
\subsection{Calculation of latencies of necessary delay blocks}
The latency of each delay block must be known at the elaboration time.
Therefore the analysis tool generates a package, implementing the function,
which accepts two parameters: the unique ID of the LCEQ block (LEQ\_ID) and the number
of the path in this block. This function returns required latency as an integer value.
The analysis tool (latreadgen.py) is written in Python. Its calling syntax is
as follows:
\begin{verbatim}
latreadgen.py /file/with_time_markers package_file package_name function_name
\end{verbatim}
An example call, as used in the demonstration project makefile is:
\begin{verbatim}
latreadgen.py /tmp/latrep.txt lateq_read_pkg.vhd lateq_read_pkg lateq_read_delays
\end{verbatim}
The analysis tools reads the time markers reported in each clock cycle,
checks if their difference remains constant during the whole simulation (except 
the initialization phase, when at least one of time markers has the initial value),
and calculates the needed additional delay as a difference between the lowest time
marker and the time marker on the input of the particular path.

An example of generated latency configuration function is shown in Figure~\ref{fig:read-lat-ex}
	\begin{figure}
	\begin{minipage}{\linewidth}
	{
	\scriptsize
	\begin{multicols}{2}
\begin{verbatim}
library ieee;
use ieee.std_logic_1164.all;
use ieee.numeric_std.all;
use std.textio.all;

library work;
use work.lateq_pkg.all;

package lateq_read_pkg is

  function lateq_read_delays (
    constant leq_id : in string;
    constant n : in integer)
    return integer;

end package lateq_read_pkg;

package body lateq_read_pkg is

  function lateq_read_delays (
    constant leq_id : in string;
    constant n : in integer)
    return integer is
  begin  -- function lateq_read_delays
    if leq_id = "" then
      return 0;
    elsif leq_id="LCEQ1" then
      case n is
        when 0 => return 4;
        when 1 => return 0;
        when others => return -1;
      end case;
    elsif leq_id="LCEQ2" then
      case n is
        when 0 => return 5;
        when 1 => return 1;
        when 2 => return 0;
        when others => return -1;
      end case;
    else
      return 0;
    end if;
  end function lateq_read_delays;

end package body lateq_read_pkg;
\end{verbatim}
	\end{multicols}
	}
	\end{minipage}
	\vspace{3mm}
	\caption{\label{fig:read-lat-ex}
	An example of the generated function returning the calculated latency of each delay block in each LCEQ block.
	The function is generated by the latreadgen.py tool from the recorded time marker reports.
	}
	\end{figure}

This approach has an additional advantage that the final sources with properly
balanced latencies (which may be used for synthesis) contain only the standard
VHDL files.
\subsection{Unique identifiers of LCEQ blocks}
Both reporting of time markers and configuration of latencies of delay blocks require
that each LCEQ block in the design has its unique identifier. It must be the same 
during the simulation and during the synthesis.
Theoretically VHDL offers the INSTANCE\_NAME attribute, which should univocally identify each instance of each component used in the design. Unfortunately, the tests have shown, that each simulation or synthesis tool may use a slightly different
format of the generated identifier. Additionally, during the simulation the system is instantiated in the testbench, while, during the synthesis, it may be either a
top entity or may be a component of a bigger system. That also leads
to different INSTANCE\_NAME values during the simulation and during the synthesis.

To work around those problems, the LCEQ block is equipped with generic LEQ\_ID of string type. This generic should be set to the unique LCEQ identifier during the instantiation
of the block.
If this block is instantiated inside of another block, then this ``container'' block
should be also equipped with its unique ID.
In such case, during instantiation of the internal LCEQ block, its  LEQ\_ID should be set to the value:

\verb|"ID_of_container_block:ID_of_LCEQ_block"|

If the block is instantiated in the for-generate loop, the loop variable should be converted to the string (using the \verb|integer'image| function), and concatenated to the ID of the instantiated block.
Such approach allows to create unique identifiers, portable between different simulation and synthesis tools.

\subsection{Generation of the dedicated LCEQ blocks for different types of data}
\label{sec:lceq-generation}
As it was mentioned in section~\ref{sec:lateqgen-1st}, if the paths analyzed and equalized by the LCEQ blocks do not use the same type of data, it is necessary to generate the source code of the specialized LCEQ block for each combination of
data types.
The demonstration implementation provides such a tool, named ``lateqgen.py'', which
 should be called with the following arguments:
\begin{itemize}
 \item Entity name of the generated block.
 \item Path to the file in which the sources of the block are to be generated.
 \item List of the types of data used in consecutive data paths of the created block. Length of this list defines the number of data paths in the block.
\end{itemize}
For example to generate the lceq1.vhd file with sources of the entity lceq1 
implementing the LCEQ block with four paths where the first two of them handle 
data of type \verb|T_VOLTAGE|, the third one uses data of type \verb|T_WIDTH| and the fourth - \verb|T_POSITION|, the user should call that tool as:

\verb|lateqgen.py lceq1 lceq1.vhd T_VOLTAGE T_VOLTAGE T_WIDTH T_POSITION|	
	
Due to the way how the code is generated there are some limitations on the names 
of the data types handled by the generated LCEQ blocks.
The name of each type should start with \verb|T_|. Additionally for each such type 
the user should define the constant providing the initial value of signals
of that type. The name of that constant must be derived from the name of the type
by replacing the initial \verb|T_| with \verb|C_| and by adding \verb|_INIT| at the end.

\section{Results}
\label{sec:example-system}
To verify the proposed methodology, the example 
data processing system has been included in the sources~\wzcite{url-opencores-lateq}.
\subsection{Test data processing system}
The system receives data from ADC converters connected to $M$ readout channels 
of a particle detector. The voltage level in each channel is proportional
to the amount of charge received by that channel in the previous clock period.
The particle passing through the detector generates a certain charge that is distributed
between neighbouring channels. The amount of this charge is proportional to the particle's energy, and the center of gravity of the collected charge defines the position of the hit.

In each clock cycle, the system finds the number of the channel with the highest level of the signal $N_{max}$. This value is treated as a non-interpolated position of the hit $X=N_{max}$.
Then the system selects signals from this channel and $K$ neighbouring channels at each side: $V_i$ for $N_{max}-K < i < N_max+K$. Next, the system calculates the sum of charges (basing on the proportionality between the charge and the voltage):
	$$S=\sum_{i=N_{max}-K}^{N_{max}+K}{V_i}$$
 and the weighted sum of charges:
	$$S_{W}=\sum_{i=N_{max}-K}^{N_{max}+K}{i \cdot V_i}$$
Calculated values are transmitted to the external system (in the simulation to the
testbench), which calculates the center of gravity of the charge and finally, the interpolated position of the particle hit:
	
$$X=N_{max}+\frac{S_{W}}{S}$$
	
\begin{figure}[t]
 {   
   \begin{center}
   \begin{tabular}{c}
	   \includegraphics[width=0.9\linewidth]{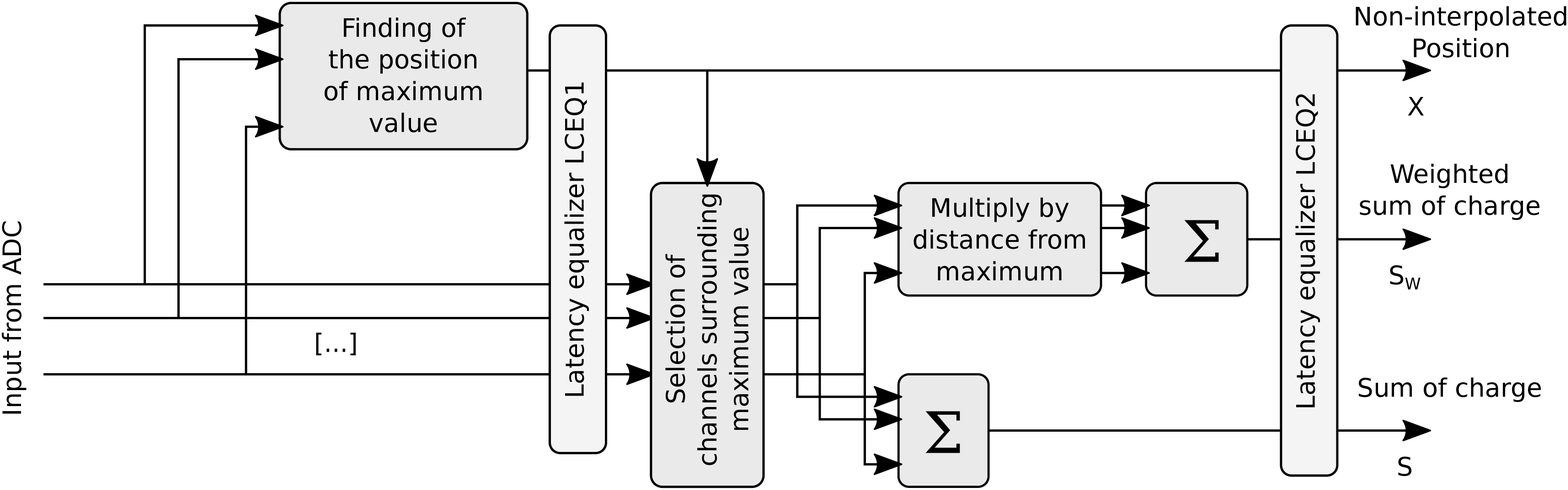}
   \end{tabular}
   \end{center}
   \caption
	{ \label{fig:sample-sys1}
	Block diagram of the example system using the same data type in all paths.
	}
   }
\end{figure}

The block diagram of the example system is shown in Figure~\ref{fig:sample-sys1}
	
Please note, that this block is not of production quality. E.g., it may incorrectly
handle the situation, where the maximum signal is too near to the edge of the detector
	(i.e. $N_max<K$ or $N_{max}>M-1-K$).
\begin{figure}[t]
 {   
   \begin{center}
   \begin{tabular}{c}
	   \includegraphics[width=0.9\linewidth]{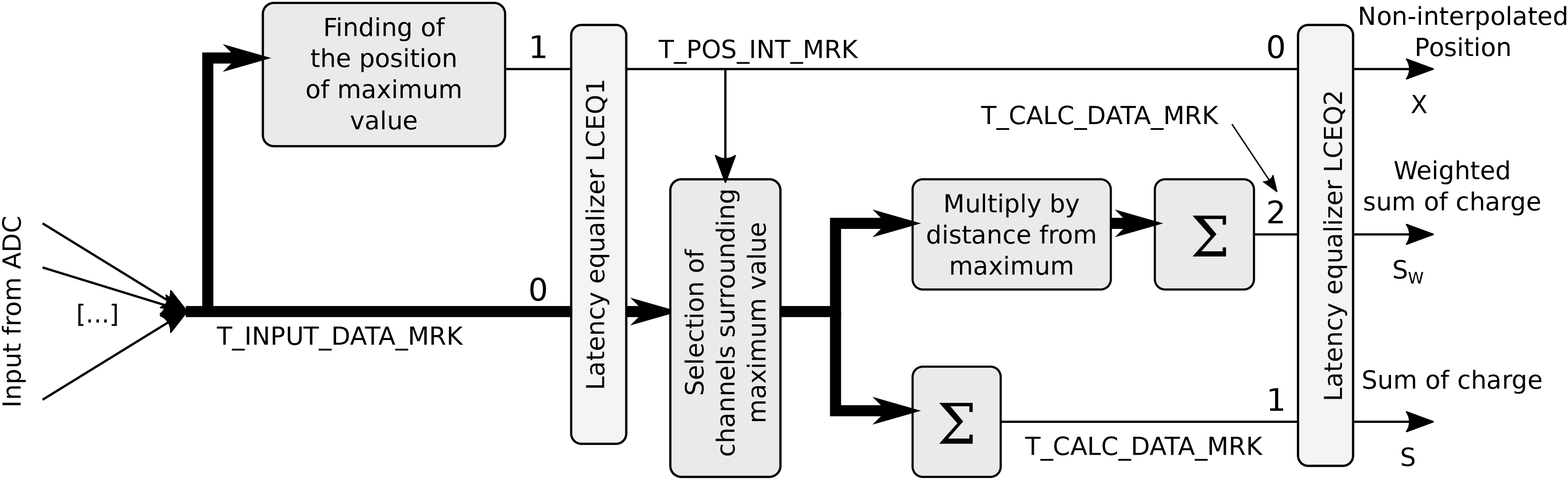}
   \end{tabular}
   \end{center}
   \caption
	{ \label{fig:sample-sys2}
	Block diagram of the example system using various types in different data paths.
	The most important type names are written in the figure.
	The path numbers are also shown at the inputs of the LCEQ blocks (they are referred to
	in the section describing the results).
	}
   }
\end{figure}

The latency of different paths in the example system may be modified by adjustment
of certain parameters in the \verb|ex1_pkg.vhd| and \verb|ex1_trees_pkg.vhd| files.

Finding the maximum value is performed in a multi-level tree based comparator consisting of a certain number 	of basic comparators. 
The number of inputs of each basic comparator should be chosen depending on the hardware features of the particular FPGA device and required speed of operation.
Each level of the comparator is also equipped with a pipeline register. 
Therefore, the total latency of the whole ``Maximum Value Finder'' block depends
on the number of inputs in the entire system (parameter \verb|C_N_CHANNELS| in
\verb|ex1_pkg.vhd|) and also on the number of inputs in a single basic comparator
(parameter \verb|EX1_NOF_INS_IN_CMP| in \verb|ex1_trees_pkg.vhd|).

Similarly the adders calculating the sum of charge and the weighted sum of charge have a multilevel tree-based structure, and again their latency depends on the number of channels selected for those calculations (parameter \verb|C_N_SIDE_CHANS| in \verb|ex1_pkg.vhd|) and the number of inputs in a basic adder 
(parameter \verb|EX1_NOF_INS_IN_ADD| in \verb|ex1_trees.pkg.vhd|).

There are two implementations of the demonstration system. The first one, located in
the \verb|hdl_single_type| directory uses one type \verb|T_USER_DATA_MRK| in all paths
in the system. That allows to avoid using generated LCEQ blocks but requires
additional effort to find a common representation for different data (the input signal,
the sum of charges, the position of maximum, etc.).
The second implementation, located in the \verb|hdl_various_types| directory,
shows how to use the proposed methodology with different types, individually 
suited for different kinds of information processed in the system. Therefore
the LCEQ blocks are generated as follows:

\verb|lateqgen.py ex1_eq_mf ex1_eq_mf.vhd T_INPUT_DATA_MRK T_POS_INT_MRK|

\verb|lateqgen.py ex1_eq_calc ex1_eq_calc.vhd T_POS_INT_MRK \|\\
\verb|                        T_CALC_DATA_MRK T_CALC_DATA_MRK|

Provided sample implementation is licensed under the BSD license, so it may be used
not only to verify and investigate proposed methodology but also as a starting point for its adoption in user's own projects.

\subsection{Tests of the proposed method}
In the described parameterized implementation of the test system, each change of its parameters
may result in a change of latency of corresponding paths.
Without the described method, these latencies should be afterwards manually balanced by the user.
Thanks to the proposed method, the user may perform automatic equalization of latencies.
During the tests, the parameters described in the previous subsection were changed, and the
additional latencies calculated by the proposed method were checked.
Correct operation of the system was also verified, using the simulated hit data in the testbench.
Obtained results are presented in Table~\ref{tab:results-latencies}.
In all cases,  the correct operation of the system after latency balancing was confirmed.
\begin{table}[tp]
  \caption[results]
  { \label{tab:results-latencies}
    The results of latency adjustments for different values of parameters of the test system.
    The upper part of the table shows the parameters values for different test cases, the lower
    part - the values of additional latencies calculated by the method. The path numbers are defined
    in sources and shown in Figure~\ref{fig:sample-sys2}.
    In all cases, the correct operation of the system after latency balancing was confirmed.
    Only results for implementation with various data types is shown, as the number of paths
    in the LCEQ1 block for single type implementation is very high.
  }
\begin{center}
{\small
\begin{tabular}{|c|c|c|c|c|c|c|c|c|c|}
\hline
\multicolumn{2}{|c}{\multirow{2}{*}{Parameter name}} &
\multicolumn{7}{|c|}{Test case}\\
\cline{3-9}
\multicolumn{2}{|c|}{}                                 & 1  & 2 & 3 & 4 & 5 & 6 & 7 \\
\hline
\multicolumn{2}{|c|}{C\_N\_CHANNELS}                   & 64 & 64 & 32 & 32  & 64 & 64 & 64 \\
\multicolumn{2}{|c|}{C\_N\_SIDE\_CHANS}                &  3 &  3 &  3 &  3  &  5 &  5 & 5  \\
\multicolumn{2}{|c|}{EX1\_NOF\_INS\_IN\_CMP}           &  3 &  3 &  2 &  2  &  2 &  3 & 3  \\
\multicolumn{2}{|c|}{EX1\_NOF\_INS\_IN\_ADD}           &  3 &  2 &  3 &  2  &  2 &  2 & 3  \\
\hline
\hline
LCEQ block          & Path   & \multicolumn{7}{c|}{Calculated additional latency}\\
\hline
\multirow{2}{*}{LCEQ1} & 0                             & 4  &  4 &  5 &  5 &  6 & 4  & 4  \\
                       & 1                             & 0  &  0 &  0 &  0 &  0 & 0  & 0  \\
\hline
\multirow{2}{*}{LCEQ2} & 0                             & 4  &  5 &  4 &  5 &  6 & 6  & 5  \\
                       & 1                             & 1  &  1 &  1 &  1 &  1 & 1  & 1  \\
                       & 2                             & 0  &  0 &  0 &  0 &  0 & 0  & 0  \\
\hline

\end{tabular}
}
\end{center}
\end{table}

To allow the user to verify the presented results, and to allow to perform experiments with
the modified or own design, the dedicated makefile is prepared.
To run the provided demonstration the user must have installed on his or her computer
Python version 3\wzcite{url-python}, GHDL simulator\wzcite{url-ghdl} and GTKWave viewer\wzcite{url-gtkwave}.
The test makefile defines a few targets:
\begin{itemize}
\item {\bf make clean} - removes the compiled files and simulation results.
\item {\bf make initial} - generates the initial version of latency configuration
function, which sets latency to 0 in all paths of all LCEQ blocks.
\item {\bf make final} - performs simulation in the ``final test'' mode. If latencies are not properly balanced, one should expect error messages about unequal latencies\\
(e.g.: \verb|EQ1 inequal latencies: out0=0, out1=-1|).
\item {\bf make synchro} - performs the simulation-analysis-correction cycle. After this command, the latencies should be properly equalized, and further running of ``make final'' should not report any errors. In fact, the testbench should also report
two correctly analyzed particle hits like below:\\
Hit with charge: 2.5e2 at 1.476e1 \\
Hit with charge: 2.65e2 at 2.549e1
\item {\bf make reader} - allows to start the GTKWave viewer and see values of the signals
in the demonstration system during the last simulation. This target may be used to analyze the internals of the system.
\end{itemize}

\subsection{Tests of synthesizability of the generated sources}
The sources generated by the test makefile with ``synchro'' target have been successfully synthesized with the Xilinx Vivado\cite{url-xlx-vivado} tools. The blocks related to time markers generation and checking have been
correctly removed from the synthesized design, and only the additional delay blocks have been inserted.
Due to high number of pins, the xc7vx690tffg1930 Virtex 7 chip was selected for implementation.
	
\section{Conclusions}
The method presented in this paper is a solution of an important problem of equalization of latencies between parallel paths in complex pipelined data processing systems implemented in FPGA.
The method extends the concept of simulation based pipeline delay balancing method offered by the
``sync'' block in the early versions of Xilinx System Generator for Simulink environment.
The solution described in the paper is suitable for systems implemented entirely in the VHDL,
and should be compatible with all recent simulation and synthesis tools.
The simulation of the designed subsystem allows to calculate latencies 
of necessary additional delay blocks in a single simulation-analysis-correction cycle.
The method allows to equalize latency between paths with data of different types
which is crucial in complex systems.
To achieve that, dedicated tools have been written in Python 3 to overcome limitations
of the VHDL language	and to generate source code of necessary blocks.
The results of latency equalization are implemented in a standard VHDL package with function defining latencies of all added delay blocks. 

The sources of the first ``proof of the concept'' implementation of the proposed methodology are published on the Open Cores website\wzcite{url-opencores-lateq}, under the BSD license.
The correctness of the method has been verified with the complete example data processing system included in the sources.
Further improvements of the proposed method should be focused on optimization of communication between the simulator and the latency analysis tool. Probably using the named sockets or VHPI interface may significantly improve the simulation and analysis speed.
Anyway even in the current state the proposed method may be a useful tool for designing
and maintenance of complex pipelined IP cores implemented in VHDL.

    %
    %
    %
    %
    
 %
 %
 %
 %
 %
\bibliography{br_align}   %
\bibliographystyle{unsrt}   %

\vspace{1cm}
~\\
\begin{tabular}{p{0.15\linewidth}p{0.8\linewidth}}
  \begin{minipage}[b]{\linewidth}
    \includegraphics[width=\linewidth]{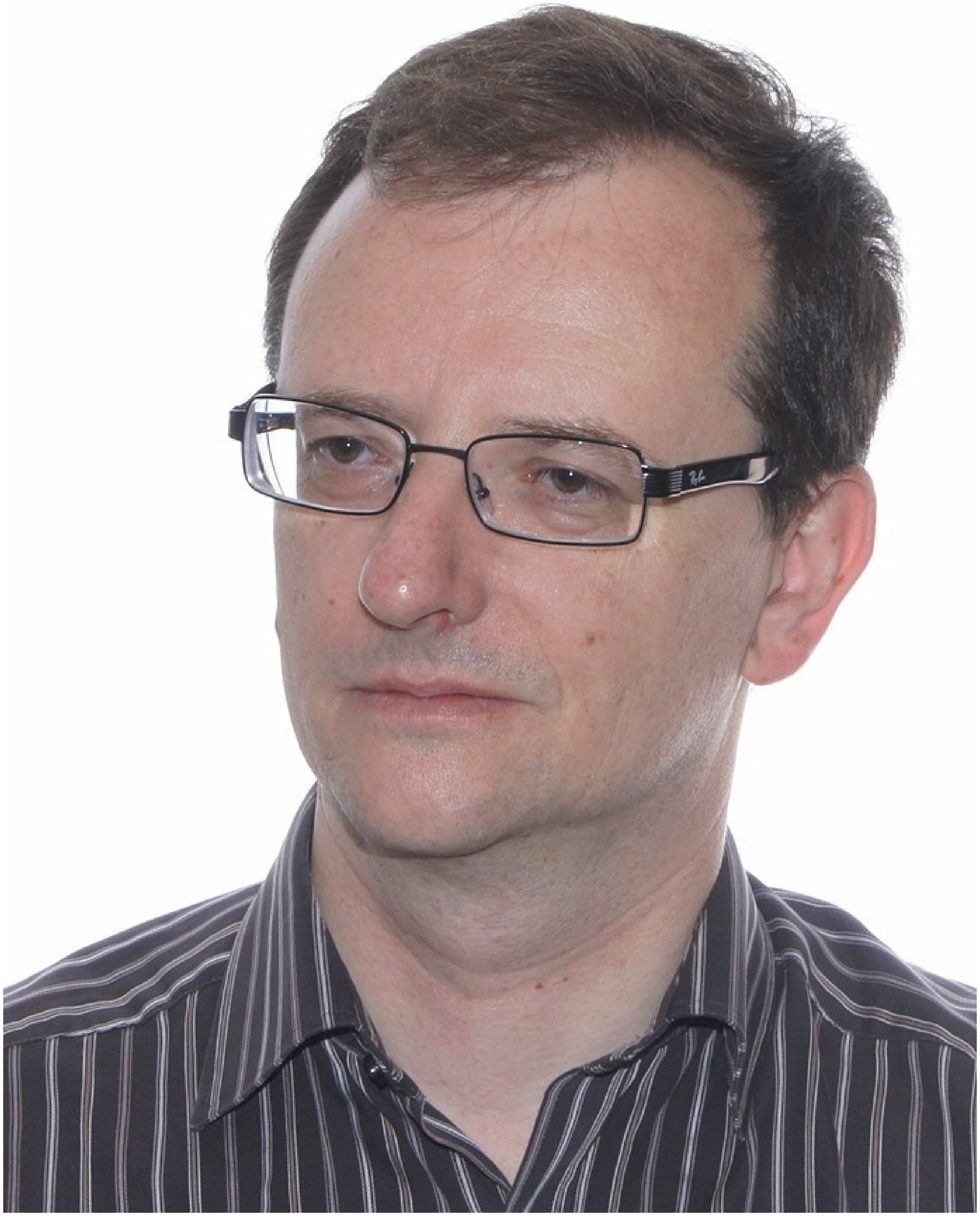} 
  \end{minipage}
&
  \begin{minipage}[b]{\linewidth}
     \input{wzab_cv}

  \end{minipage}
\\
\end{tabular}

\end{document}

%% file: wz_commands.tex
\newcommand{\wzcite}[1]{ \cite{#1}}

%% file: wzab_cv.tex
Wojciech M. Zabolotny was born in Sucha Beskidzka,
Poland in 1966. He received the MSc (1989)
and the Ph.D. (1999) in Electronics from the
Warsaw University of Technology in Poland, both with honors.
Since 1990 he was a research assistant and since 1999 he
is an Assistant Professor at the 
Warsaw University of Technology.
His research interests are the distributed data acquisition
systems (biomedical and for high energy physics),
the embedded systems and programmable logic.
He was involved in development of electronic systems for
CERN (since 2002), for DESY in Hamburg (2002-2009),
for CBM in Darmstadt (since 2008),
and for JET in Culham (since 2010).